\documentclass[english,aps,pre,superscriptaddress,twocolumn,showpacs]{revtex4}
\usepackage{graphicx,amsfonts,amsmath,amssymb}
\usepackage[english]{babel}

\newcommand{\p}[2]{\frac{\partial\, #1}{\partial\, #2}\,}
\newcommand{\e}{\varepsilon}
\newcommand{\Z}{\mathbb{Z}}
\newcommand{\te}[1]{\,\mbox{#1}\,}
\newcommand{\V}[1]{\mathbf{#1}}
\newcommand{\intl}[2]{\int\limits_{#1}^{#2}}
\newcommand{\bracket}[1]{\left(#1\right)}

\newcommand{\eq}[1]{$\mathrm{Eq.}$~\eqref{#1}}
\newcommand{\figref}[1]{$\mathrm{Fig.}$~\ref{#1}}
\newcommand{\secref}[1]{$\mathrm{Sec.}$~\ref{#1}}

\newcommand{\av}[1]{\left\langle #1 \right\rangle}

\DeclareMathOperator{\asinh}{asinh}

\begin{document}

\title{Entropic particle transport: higher order corrections to the Fick-Jacobs diffusion equation}

\author{S. Martens}
\email{steffen.martens@physik.hu-berlin.de}
\affiliation{Department of Physics, Humboldt-Universit\"{a}t zu Berlin, Newtonstr. 15, 12489 Berlin, Germany}
\author{G. Schmid} 
\affiliation{Department of Physics, Universit\"{a}t Augsburg, Universit\"{a}tsstr. 1, 86135 Augsburg, Germany}
\author{L. Schimansky-Geier}
\affiliation{Department of Physics, Humboldt-Universit\"{a}t zu Berlin, Newtonstr. 15, 12489 Berlin, Germany}
\author{P. H\"anggi}
\affiliation{Department of Physics, Universit\"{a}t Augsburg, Universit\"{a}tsstr. 1, 86135 Augsburg, Germany}

\begin{abstract}
\noindent Transport of point-size Brownian particles under the
influence of a constant and uniform force field through a
three-dimensional  channel with smoothly varying periodic
cross-section is investigated. Here, we employ an asymptotic
analysis in the ratio between the difference of the widest and the most narrow
constriction divided through the period length of the channel
geometry. We demonstrate that the leading order term is equivalent
to the Fick-Jacobs approximation. By use of the higher order
corrections to the probability density we derive an expression for
the spatially dependent diffusion coefficient $D(x)$ which
substitutes the constant diffusion coefficient present in the common
Fick-Jacobs equation. In addition, we show that in the diffusion
dominated regime the average transport velocity is obtained as the
product of the zeroth-order Fick-Jacobs result and the expectation
value of the spatially dependent diffusion coefficient $\av{D(x)}$.
The analytic findings are corroborated with the precise numerical
results of a finite element calculation of the Smoluchowski
diffusive particle dynamics occurring in a reflection symmetric
sinusoidal-shaped channel.
\end{abstract}

\maketitle

\section{Introduction}

The transport of large molecules and small particles that are
geometrically confined within pores, channels or other
quasi-one-dimensional systems attracted attention in the last
decade. This activity stems from the  profitableness for shape and
size selective catalysis, particle separation and the dynamical
characterization of polymers during their translocation
\cite{Burada2009_CPC,Hanggi2009,Keil2000,Kettner2000,Dekker2007}. In
particular, the latter theme which aims at the experimental
determination of the structural properties and the amino acid
sequence in DNA or RNA when they pass through narrow openings or the
so-called bottlenecks, comprises challenges for technical
developments of nanoscaled channel structures
\cite{Dekker2007,Keyser2006,Howorka2009,Pedone2010}.

Along with the progress of the experimental techniques the problem
of particle transport through corrugated channel structures
containing narrow openings and bottlenecks has give rise to recent
theoretical activities to study diffusion dynamics occurring in such
geometries \cite{Burada2009_CPC}. Previous studies by Jacobs
\cite{Jacobs} and Zwanzig \cite{Zwanzig1992} ignited a revival of
doing research in this topic. The so-called {\it Fick-Jacobs
approach} \cite{Jacobs, Zwanzig1992}, accounts for the elimination
of  transverse stochastic degrees of freedom by assuming a fast
equilibration in those transverse directions \cite{Jacobs}. The
theme found its application for particle transport  through periodic
channel structures \cite{Reguera2006} and designed single nanopores
\cite{Berezhkovskii2007} exhibiting smoothly varying side walls.
Several aspects of driven motion in presence of applied external
force fields and the quality of  the Fick-Jacobs approach in
presence of an applied force field in corrugated structures has been
the focus of recent studies \cite{Burada2007, Burada2008,
Kalinay2006, Bradley2009, Kalinay2009, riefler2010}.

Beyond the Fick-Jacobs (FJ) approach, which is suitably applied to
channel geometries with smoothly varying side walls, there exist yet
other methods for describing the transport through varying channel
structures like cylindrical septate channels
\cite{Marchesoni2010,Borromeo2010,Hanggi2010}, tubes formed by
spherical compartments \cite{Berezhkovskii2010a, Berezhkovskii2010b}
or channels containing  abrupt changes of cross diameters
\cite{Kalinay2010}.

Our objective with this work is to provide a systematic treatment by
using a series expansion in terms of a smallness parameter which
specifies the channel corrugation for {\it biased} particle
transport proceeding along an extended, three-dimensional periodic,
reflection symmetric channel for which the original, commonly
employed (lowest order) Fick-Jacobs approach fails because of
extreme bending of the channel's side walls.

In \secref{sec:structure} we introduce the model system: a Brownian
particle in a confined channel geometry with reflection symmetric,
irregular boundaries. The central findings, namely the analytic
expressions for the probability density and the average transport
characteristics  are presented in \secref{sect:longwave}. In
\secref{sec:part_sinuschannel} we employ our  analytical results to
a specific channel configuration consisting of sinusoidally varying
side walls. Section~\ref{sec:conclusion} summarizes our findings.

\section{Transport in confined structures}
\label{sec:structure}

\begin{figure}[t]
  \centering
  \includegraphics{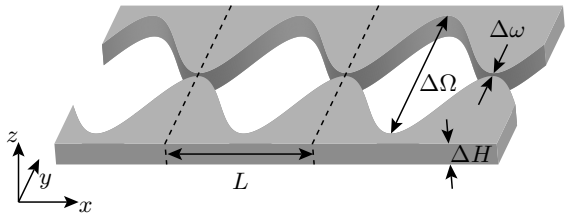}
  \caption{Sketch of a segment of a reflection-symmetric sinusoidally varying channel that is
    confining the motion of the overdamped, point-like Brownian particle. The periodicity of the channel structures is $L$, the height $\Delta H$, the minimal and
    maximal channel widths are $\Delta \omega$ and $\Delta \Omega$,
    respectively. The size of an unit-cell is indicated with the dashed lines.}
  \label{fig:channelprob}
\end{figure}

Generic mass transport through confined structures such as irregular
pores and channels occurs due to the combination of molecular
diffusion, as quantified by the molecular diffusivity $D$, and
passive transport arising either from different particle
concentrations maintained at the ends of the channel, an applied
hydrodynamic velocity field or an external, force generating
potential $U(x,y,z)$. Here, we concentrate on constant force-driven
transport where particles of dilute concentration (i.e. interaction
effects can safely be neglected)  are subjected to a fixed external
force with magnitude $F$ acting along the longitudinal direction of
the channel $\V{e}_x$, i.e., $U(x,y,z)=-F\,x$. The overdamped single Brownian
particle then budges in a three-dimensional periodic channel
geometry of period $L$, constant height $\Delta H$, and periodically
varying transverse width. A sketch of a segment of the channel is
depicted in \figref{fig:channelprob}. The shape of the side walls are
described by the two boundary functions $\omega_\pm(x)$. As we
restrict ourselves to reflection-symmetric confinements in
$y$-direction, we set $\omega_\pm(x) \equiv \pm \omega(x)$.

 The evolution of the probability density $P\bracket{\V{q},t}$ of
 finding the particle at the local position $\V{q}=\bracket{x,y,z}^T$ at time $t$ is governed by the
 three-dimensional Smoluchowski equation
 \cite{Risken,Haenggi1982}, i.e.,
 \begin{subequations}
   \label{eq:sm}
   \begin{align}
     &\partial_t P\bracket{\V{q},t}+\nabla_{\V{q}} \cdot \V{J}\bracket{\V{q},t}=\,0\,, \label{eq:Smoluchowski}\\
     \intertext{where}
     &\V{J}\bracket{\V{q},t}=\,\frac{F}{\eta}\,P\bracket{\V{q},t}\,\V{e}_x-\frac{k_BT}
     {\eta}\,\nabla_{\V{q}}\,P\bracket{\V{q},t}\, \label{eq:def_probcurr}
     \end{align}
 \end{subequations}
is the probability current of the probability density
$P\bracket{\V{q},t}$. The force strength acting on the Brownian
particle is denoted by $F$, $\eta$ is the friction coefficient,
while the Boltzmann constant is $k_B$ and $T$ refers to the
environmental temperature. Because of the impenetrability of the
channel walls the probability current
$\V{J}\bracket{\V{q},t}=\bracket{J^x,J^y,J^z}^T$ is subjected to the
no-flux boundary condition, reading
\begin{align}
 \V{J}\bracket{\V{q},t}\cdot\V{n}=0\,,\quad \forall \V{q} \in \mbox{channel wall}\,. \label{eq:bc}
\end{align}
$\V{n}$ denotes the out-pointing normal vector at the channel walls. 
The probability density satisfies the normalization condition
$\int_{\mathrm{unit-cell}} P(\V{q},t)\, d^3\V{q} =1$.
as well as the periodicity condition $P(x+m\,L,y,z,t)=P(x,y,z,t)\,,
\forall m \in \Z$. In the long time limit the stationary probability density is defined as
$P_{\mathrm{st}}\bracket{\V{q}}:=\lim_{t\to \infty}
P\bracket{\V{q},t}$. Analogously, the stationary probability current reads
$\V{J}_{\mathrm{st}}\bracket{\V{q}} :=\lim_{t\to \infty}
\V{J}\bracket{\V{q},t}$.

The  key quantities of particle transport through such periodic
channels are the average particle velocity $\av{\V{\dot{q}}}$ and
the effective diffusivity $D_\mathrm{eff}$. The latter is given by
\begin{align}
 D_\mathrm{eff}=\lim_{t \to \infty} \frac{\av{x^2(t)}-\av{x(t)}^2}{2 t}\,, \label{eq:Deff}
\end{align}
and can be calculated by means of the stationary probability density
$P_{\mathrm{st}}\bracket{\V{q}}$ using an established method taken
from Ref.~\cite{Brenner}. Once $P_{\mathrm{st}}(\V{q})$ is known, the
mean particle velocity of Brownian particles can be computed by
\begin{align}
  \av{\V{\dot{q}}}\equiv&\,\lim_{t\to\infty} \frac{\av{\V{q}(t)}}{t} =
  \intl{\mathrm{unit- cell}}{}\,\V{J}_{\mathrm{st}}(\V{q})\,
  d^3\V{q} \label{eq:velocity}\,.
\end{align}

We next introduce dimensionless variables. In doing so, we measure
longitudinal length and height as $\overline{x}=x/L$ and
$\overline{z}=z/L$, respectively. For the rescaling of the
$y$-coordinate, we introduce the dimensionless aspect parameter $\e$: This is the difference of the widest cross-section
of the channel, i.e. $\Delta\Omega$, and the most narrow constriction at the
bottleneck, i.e. $\Delta \omega$, in units of the period length,
yielding
\begin{align}
 \e=\frac{\bracket{\Delta\Omega-\Delta\omega}}{L}\,. \label{eq:def_eps}
\end{align}
The dimensionless value of $\e$ characterizes the deviation of the
boundary from the straight channel which amounts to $\e=0$. 
Following the reasoning in Ref.~\cite{Laachi2007}, we next measure,
for the case of finite corrugation 
$\varepsilon \neq 0$, the transverse length
$y$ in units of $\varepsilon L$, i.e. $y =\varepsilon
L\,\overline{y}$ and, likewise, the boundary functions $h_\pm(x) = \omega_\pm(x)/
(\e L)$. Time is measured in units of $\tau=\, L^2 \eta /(k_B\,T)$
which is twice the time the particle assumes to overcome
diffusively, at zero bias $F=0$,  the distance $L$, i.e.
$\overline{t} = t / \tau$. The potential energy is rescaled by the
thermal energy $k_{\mathrm{B} } T$, i.e., for the considered
situation with a constant force component in channel direction:
$\overline{U} = - F x / (k_{\mathrm{B}} T) = -f \overline{x}$, with
the dimensionless force magnitude \cite{Reguera2006, Burada2008}:
\begin{align}
 f=\frac{F\,L}{k_B\,T}\,.
\end{align}
The dimensionless forcing parameter $f$ is given as the ratio of the
work $F\,L$ done on the particle when dragged by the constant force
$F$ along a distance of the period length $L$ divided by the thermal
energy $k_B T$. Note, that for an adjustment of a certain value of
$f$ in an experimental setup one can modify either the force
strength $F$ or the temperature $T$. After scaling the probability distribution reads
$\overline{P}\bracket{\V{\overline{q}},\overline{t}}=\e\,L^3\,P\bracket{\V{q},t}$, respectively, 
the probability current is given by $\V{\overline{J}}\bracket{\V{\overline{q}},\overline{t}}=\,\tau\,L^2 \bracket{\e J^x,J^y,\e J^z}^T$.
In the following, we shall omit the overbar in our notation.

In dimensionless units, the Smoluchowski equation, cf. Eqs.~\eqref{eq:sm}, reads:
\begin{subequations}
  \label{eq:sm_scaled}
  \begin{align}
    &\partial_t P\bracket{\V{q},t}+\nabla_{\V{q}} \cdot
    \V{J}\bracket{\V{q},t}=\,0\, , \label{eq:Smoluchowski_scaled}\\
    \intertext{where $\nabla_{\V{q}}=\bracket{\partial_x,\frac{1}{\e}\partial_y, \partial_z}^T$ and}
    &\V{J}\bracket{\V{q},t}=\,f \, P\bracket{\V{q},t}\,\V{e}_x- \nabla_{\V{q}}\,P\bracket{\V{q},t}\,. \label{eq:def_probcurr_scaled}
  \end{align}
\end{subequations}
At steady state, \eq{eq:Smoluchowski_scaled} becomes:
\begin{align}
  \varepsilon^2\partial_xJ_\mathrm{st}^x +\partial_yJ_\mathrm{st}^y+\varepsilon^2\partial_zJ_\mathrm{st}^z=\,0\,. \label{eq:Smoluchowski2}
\end{align}
Because (i) the dynamics in $z$-direction is decoupled from the dynamics
in $x$ and $y$-direction and (ii) the shape of the lower and upper
boundary depends neither on $x$ nor on $y$, the separation ansatz
$P_{st}(x,y,z)=\,p_{\mathrm{st}}(x,y)\,\zeta(z)$ and the boundary
condition
\begin{align}
  \label{eq:bc_z}
  J_{\mathrm{st}}^{z}=&0\,, \quad \text{at}\, z=0\, \text{and}\,z=\Delta H/L\, ,
\end{align}
results in a non-trivial solution for $\zeta(z)$ for
$J_{\mathrm{st}}^{z}\bracket{\V{q}} = 0$ everywhere within the
channel. For the considered situation, i.e. there is only a constant
force acting in $x$-direction, the form function $\zeta(z)$ equals
the inverse of the dimensionless channel height, i.e. $\zeta =
L/\Delta H$. Note, that the presented separation technique can also
be applied for more complex forcing scenarios. Assuming a general
potential landscape $U(x,y,z) = V(x,y) + W(z)$ defined within the
channel, the used separation ansatz for the stationary solution
results in
\begin{align}
P_{\mathrm{st}}(x,y,z)  = p_{\mathrm{st}}(x,y) \cdot \frac{e^{-W(z)}}{\intl{0}{\Delta H/L}dz\,e^{-W(z)}}\,.
\end{align}
Consequently, this allows a reduction of the problem's dimensionality from 3D to 2D:
\begin{align}
  \e^2 \partial_xJ_{\mathrm{st}}^{x}+&\partial_yJ_{\mathrm{st}}^{y}=0\,. \label{eq:pdf_xy}
\end{align}
Note, that the 2D transport problem was investigated in symmetric
\cite{Laachi2007,Marchesoni2009,Kalinay2006,Reguera2006,Burada2007,Burada2008,Burada2010}
and asymmetric \cite{Bradley2009,Kalinay2010} channels. For an
arbitrary dimensionless channel geometry $h_\pm\bracket{x}$ the
outwards pointing normal vector at the perpendicular side walls is
given by $\V{n}=  \bracket{ \mp h_\pm^{'}(x), \pm 1,0}^T /
\sqrt{1+h_\pm^{'}(x)^2}$ with the prime denoting the differentiation
with respect to $x$. Therefore, the no-flux boundary condition \eq{eq:bc} can
be written as
\begin{align}
  \varepsilon^2 h_\pm^{'}(x)\,J_{\mathrm{st}}^{x}=&J_{\mathrm{st}}^{y}\,, \quad \forall y \in h_\pm(x)\,.\label{eq:bc2}
 \end{align}
Note that even in the case of a more general substrate potential given
by $U(\V{q})=V(x,y)+W(z)$ the
$2$D problem \eq{eq:pdf_xy} does not dependent on the potential $W(z)$.

Finally, we define the marginal one-dimensional probability density in force direction $p_{\mathrm{st}}(x)$ as follows
\begin{align}
 p_{\mathrm{st}}\bracket{x}=\intl{h_-(x)}{h_+(x)}dy\,\intl{0}{\Delta H/L}dz\,P_{\mathrm{st}}(x,y,z)\,. \label{eq:proj_p}
\end{align}

\section{Asymptotic analysis}
\label{sect:longwave}

We apply the  asymptotic analysis
\cite{Yariv2007,Laachi2007,Wang2009} to the problem stated by
\eq{eq:pdf_xy} and \eq{eq:bc2}. In doing so, we use for the
stationary probability density $p_{st}(x,y)$ (the index $st$ will be
omitted in the following) the ansatz
\begin{align}
 p(x,y)=\sum_{n=0}^{\infty} \varepsilon^{2n} p_n(x,y)\,, \label{eq:prob_series}
\end{align}
 and for  the probability flux
\begin{align}
 \V{J}(x,y)=\sum_{n=0}^{\infty} \varepsilon^{2n} \V{J}_n(x,y)
\end{align}
in the form of a formal perturbation series in even orders of the
parameter $\varepsilon$. Substituting these expressions into
\eq{eq:pdf_xy}, we find
\begin{subequations}
\label{eq:longwave}
\begin{align}
0=&\,\partial_y J_0^y(x,y)+\sum_{n=1}^\infty \varepsilon^{2n}
\left\{\partial_x J_{n-1}^x (x,y)+\partial_y
  J_n^y(x,y)\right\}\,, \label{eq:longwave_pn}
\intertext{and the no-flux boundary condition  at the channel walls
  \eq{eq:bc2} turns into}
0=&\,-J_0^y(x,y)+\sum_{n=1}^\infty \varepsilon^{2n}\left\{h_\pm^{'}(x)J_{n-1}^x(x,y)-J_n^y(x,y)\right\}\,. \label{eq:longwave_bc}
\end{align}
\end{subequations}
Each order $p_n$ has to obey the periodic boundary condition $p_n(x+m,y)=p_n(x,y)\,,\,\forall m \in \Z$ and $p(x,y)$
has to be normalized for every value of $\e$.

Consequently, the average particle velocity  is given by
\begin{align}
 \av{\dot{x}}=&\av{\dot{x}}_0+\sum_{n=1}^\infty\e^{2n}\left\{f\av{p_n(x,y)}_{x,y}-\av{\partial_x p_n(x,y)}_{x,y}\right\}\,. \label{eq:series_vx}
\end{align}
In \eq{eq:series_vx}, the average of an arbitrary function $k(x,y)$
is defined as the integral over the cross-section in $y$ and over
one period divided by the period length which is one in the
considered scaling, i.e.
$\av{k(x,y)}_{x,y}=\int_{0}^{1}dx\int_{-h(x)}^{h(x)}dy\, k(x,y)$. In
\secref{subsect:FJ}, we demonstrate that the zeroth order of the
perturbation series expansion coincides with the Fick-Jacobs
equation \cite{Jacobs,Zwanzig1992}. Referring to \cite{Stratonovich,
  Reguera2006} an expression for the average velocity $\av{\dot{x}}_0$ is known. Moreover, in \secref{subsect:higherorders}, the higher
orders of the probability density are derived. Using those results
we are able to obtain corrections, see in \secref{subsect:spatD}, to
the average velocity beyond the zeroth order Fick-Jacobs approximation
presented in the next section.

\subsection{Zeroth Order:  the Fick-Jacobs equation}
\label{subsect:FJ}

For the zeroth order, Eqs.~\eqref{eq:longwave} read
\begin{subequations}
\begin{align}
 \partial_y J_0^{y}(x,y)=-\partial_y e^{-V(x,y)}& \partial_y\bracket{e^{V(x,y)}\,p_0(x,y)}=0\,,
\intertext{supplemented with the corresponding no-flux boundary
condition}
 J_0^y(x,y)=&\,0\,,\,\forall y \in \mbox{wall}\,.
\end{align}
\end{subequations}
Consequently,
\begin{align}
p_0(x,y)=\,g(x)\,e^{-V(x,y)}\,,
\end{align}
where $g(x)$ is an unknown function which has to be determined from
the second order $O\bracket{\varepsilon^2}$ balance given by
\eq{eq:longwave_pn}. Integrating the latter over the cross-section in
$y$ and taking the no-flux boundary conditions \eq{eq:longwave_bc}
into account, one obtains
\begin{align}
 0=&\,\partial_x \bracket{e^{-A(x)} g'(x)}\,, \label{eq:def_g}
\end{align}
where the effective potential $A(x)$ is explicitly  given by
\begin{align}
 e^{-A(x)}=\,\intl{-h(x)}{+h(x)} dy\,e^{-V(x,y)}\,. \label{eq:def_effpot}
\end{align}
For the problem at hand, i.e. for $V(x,y)=-f\,x$, as well for potentials where
$x$ enters only linearly and where $x$ is not multiplicatively coupled to the other
spatial coordinates \cite{Burada2008_PRL,Burada2009,Burada2010} the stationary
probability density within the zeroth order reads
\begin{align}
 p_0(x,y)=&e^{-V(x,y)}g(x)=\frac{e^{-V(x,y)}\intl{x}{x+1}
   e^{A(x')}dx'}{\intl{0}{1}dx e^{-A(x)}\intl{x}{x+1}
   e^{A(x')}dx'}\,. \label{eq:finalsol_p0}
\end{align}
In addition, the marginal probability density \eq{eq:proj_p} becomes
\begin{align}
p_0(x)=&\,e^{-A(x)}\,g(x)\,.  \label{eq:finalsol_projp0}
\end{align}
Expressing next $g(x)$ by $p_0(x)$, see \eq{eq:def_g}, then yields
the celebrated stationary Fick-Jacobs equation
\begin{align}
 0=\partial_x\bracket{e^{-A(x)}\partial_x e^{A(x)}\,p_0(x)} \label{eq:FJ_pdf}
\end{align}
derived  previously in Ref.~\cite{Jacobs,Zwanzig1992,Reguera2001}. Thus,
we find the result that the leading order term of the asymptotic
analysis is equivalent to the FJ-equation. Please note that the
differential equation determining the unknown function $g(x)$, cf.
\eq{eq:def_g}, is the same for the dynamics of a Brownian particle
evolving in an energetic potential $V_\mathrm{en}(x,y)$ leading to a
confinement in $y$-direction, with the natural boundary conditions
$J_n^y(x,y=\pm\infty)=0$ \cite{Zwanzig1992,Wang2009}.
Therefore, in zeroth order and for the given scaling, an appropriately chosen confining
energetic potential $V_\mathrm{en}(x,y)$ obeying
$\int_{-\infty}^{\infty} dy \exp( -V_\mathrm{en}(x,y)) =
\int_{-h(x)}^{h(x)} dy \exp( -V(x,y))$ results in the same transport
characteristics as those induced by the confining channel with the
boundary functions $h_{\pm}(x)$ \cite{Sokolov2010}.

The average particle current is calculated by integrating the probability flux
$J_0^x$ over the unit-cell \cite{Stratonovich,Schimansky1}
\begin{align}
 \av{\dot{x}(f)}_0=&\,\intl{0}{1}dx\,\intl{h_-(x)}{h_+(x)}dy\,J_0^x(x,y)
  \nonumber \\
=&\,\frac{1-e^{-f}}{\intl{0}{1}dx\,e^{-A(x)}\,\intl{x}{x+1}
  e^{A(x')}\,dx'}\,. \label{eq:currFJ}
\end{align}
In the spirit of linear response theory, the mobility in units of
the free mobility $1/\eta$ is defined by the ratio of the mean
particle current \eq{eq:currFJ} and the applied force $f$ yielding
\begin{align}
 \eta\,\mu_0\bracket{f}=\frac{\av{\dot{x}(f)}_0}{f}\,. \label{eq:def_mob}
\end{align}

\subsection{Higher order contributions to the Fick-Jacob equation}
\label{subsect:higherorders}

We next address the higher order corrections $p_n(x,y)$ of the
probability density. According to \eq{eq:longwave_pn}, one needs to
iteratively solve
\begin{align}
\label{eq:iterative}
 \partial_y^2 p_n(x,y)=&\,\mathfrak{L}\,p_{n-1}(x,y)\,,\quad n\geq 1\,,
\end{align}
under consideration of the boundary condition \eq{eq:longwave_bc}.
In \eq{eq:iterative}, we make use of the operator $\mathfrak{L}$,
reading $\mathfrak{L}=\bracket{f\,\partial_x-\partial_x^2}$. Applied
$n$-times yields the expression
\begin{align}
\mathfrak{L}^{n}=\,\sum_{k=0}^{n} \begin{pmatrix} n \\ k \end{pmatrix}
\bracket{-1}^k f^{n-k}\,\p{ ^{n+k}}{x^{n+k}}. \label{eq:Ln}
\end{align}
Each solution of the second order partial differential equation
\eq{eq:iterative} possesses two integration constants $d_{n,1}$ and
$d_{n,2}$. The first one, $d_{n,1}$, is determined by the no-flux
boundary condition \eq{eq:longwave_bc} while the second provides
 the normalization condition $\av{p(x,y)}_{x,y}=1$. In what
follows, we use the normalization constant of the probability
density $p(x,y)$ via  the zeroth order $\av{p_0(x,y)}$. As a
consequence, we have
\begin{subequations}
\begin{align}
 \av{p_0(x,y)}_{x,y}=&\intl{0}{1}dx\,\intl{h_-(x)}{h_+(x)}dy\,p_0(x,y)=\,1\,,\\
 \av{p_n(x,y)}_{x,y}=&\,0\,,\quad \forall n\geq 1\,, \label{eq:condnorm_pn}
\intertext{with the constraint that}
&\,\intl{h_-(x)}{h_+(x)}dy\,p_n(x,y)\neq 0 \,,\quad \forall n\geq
1\,,
\end{align}
\end{subequations}
in order to prevent that the marginal probability density
\eq{eq:proj_p} equals the FJ results, cf. \eq{eq:finalsol_projp0},
for an arbitrary value of $\e$, i.e. $p(x)=p_0(x)$. Further, we have
to emphasize that the centered functions
\begin{align}
{p_n}(x,y) \longmapsto \frac{p_n\bracket{x,y}-\av{p_n\bracket{x,y}}}{\av{p\bracket{x,y}}}\,,
\quad  \te{for} n\geq 1\,, \label{eq:def_shiftpn}
\end{align}
are no probability densities anymore because they can assume
negative values for a given $x$ and $y$. The calculation of the
average particle velocity \eq{eq:series_vx} simplifies to
\begin{align}
 \av{\dot{x}}=\,\av{\dot{x}}_0-\sum_{n=0}^{\infty}
 \e^{2n}\av{\partial_x p_n(x,y)}_{x,y}\,. \label{eq:series_vx_simp}
\end{align}
We find that the average particle current is composed of (i) the
Fick-Jacob result $\av{\dot{x}}_0$ , cf. \eq{eq:currFJ}, and (ii) becomes
corrected by the sum of the averaged derivatives of the higher
orders $p_n(x,y)$. One immediately notices that the second
integration constant $d_{n,2}$ does not influence the result for the
average particle velocity \eq{eq:series_vx_simp}.

For the first order correction, the determining equation is
\begin{align}
 \partial_y^2
 p_1(x,y)=&\,\mathfrak{L}\,p_{0}(x,y)=\,\frac{\av{\dot{x}}_0}{2}\,\partial_x
 \bracket{\frac{1}{h(x)}}\,,
\intertext{and after integrating twice over $y$, we obtain}
 p_1(x,y)=&\,-\frac{\av{\dot{x}}_0}{2}\,\bracket{\frac{h'(x)}{h^2(x)}}\,\frac{y^2}{2!}\,. \label{eq:finalsol_p1}
\end{align}
Hereby, as previously requested above, the first integration
constant $d_{1,1}(x)$ is set to $0$ in order to fulfill the no-flux
boundary condition, and the second must provide the normalization
condition \eq{eq:condnorm_pn}, i.e. $d_{1,2}=0$. Consequently, the
first correction to the probability density becomes positive if the
confinement is constricting, i.e. for $h'(x) <0$ and
$\av{\dot{x}}_0\neq 0$. In contrast, the probability density becomes
less in unbolting regions of the confinement, i.e. for $h'(x)>0$.
Please note, that the first order correction scales linearly with
the average particle current $\av{\dot{x}}_0$. Overall, the break of
spatial symmetry observed within numerical simulations in previous
works \cite{Burada2007,Burada2010_PRE} is reproduced by this very
first order correction. Particularly, with increasing forcing, the
probability for finding a particle close to the constricting part of
the confinement increases, cf.
Ref.~\cite{Burada2007,Burada2010_PRE}.

Upon recursively solving, we obtain for the higher order corrections
$n\geq 1$ as
\begin{align}
\begin{split}
 p_n\bracket{x,y}=&\,\mathfrak{L}^{n}
 p_{0}(x,y)\,\frac{y^{2n}}{2n!}+d_{n,2}+\\
&+\sum_{k=1}^{n} \mathfrak{L}^{n-k} d_{k,1}(x)
\frac{|y|^{2\bracket{n-k}+1}}{\bracket{2\bracket{n-k}+1}!}\,,
\end{split} \label{eq:finalsol_pn}
\end{align}
with the integration constants for the $n$-th order
\begin{subequations}
\begin{align}
&d_{n,1}(x)=\,-\partial_x\bracket{\intl{0}{h(x)} dy\,J_{n-1}^x\bracket{x,y}}\,, \\
\begin{split}
&d_{n,2}=-\bracket{\intl{0}{1} dx\,\sum_{k=1}^{n}
  \mathfrak{L}^{n-k} d_{k,1}\frac{h^{2\bracket{n-k}+2}}{\bracket{2\bracket{n-k}+2}!}\right. \\
& \left. +\intl{0}{1} dx\,\mathfrak{L}^{n}
  p_{0}(x,y)\,\frac{h^{2n+1}}{(2n+1)!}} \Big/ \intl{0}{1}
dx\,h(x)\,. \end{split}\label{eq:dn2}
\end{align}
\end{subequations}
As expected, for a reflection symmetric channel in $y$-direction
each order $p_n(x,y)$ results as well in a reflection symmetric
function. The latter consists of a term proportional to even powers
in $y$ and in addition of a sum of odd powers of $|y|$, caused by
the no-flux boundary conditions. Since each integration constant
$d_{n,1}(x)$ with $n>1$ is determined by the probability current of
the previous order, every order $p_n(x,y)$ is proportional to the
average current of the zeroth order $\av{\dot{x}}_0$. Consequently,
the $2$D probability density equals the zeroth order
$p(x,y)=p_0(x,y)=const$ for all values of $\e$ in  absence of an
external force $f=0$. Further, it follows that the average particle
current \eq{eq:series_vx_simp} scales with the average particle
current obtained from the Fick-Jacobs formalism $\av{\dot{x}}_0$ for
all values of $\e$.

\subsection{Spatially diffusion coefficient $D(x)$}
\label{subsect:spatD}

With \secref{subsect:FJ}, we could show that the dynamics of
Brownian particles in confined structures can be described
approximatively by the FJ-equation, cf. \eq{eq:FJ_pdf}. Zwanzig
\cite{Zwanzig1992} obtained this $1$D equation from the full $2$D
Smoluchowski equation upon eliminating the transverse degree of
freedom. This approximation neglects the influence of relaxation
dynamics in transverse direction, supposing that it is infinitely
fast. In a more detailed view, we have to notice that diffusing
particles pile up, or miss, at the curved wall if the channel is
getting narrower or wider as they can flow out from/ or towards the
wall in $y$ direction only at finite time. These effects are
described by the higher expansion orders $p_n(x,y)$ presented in
\eq{eq:finalsol_pn}. In the following, we aim at deriving a
dynamical equation of Smoluchowski-type, but with a diffusion
coefficient that depends on the longitudinal channel coordinate $x$.

The concept of a spatially dependent diffusion coefficient $D(x)$
was introduced by Zwanzig \cite{Zwanzig1992} and subsequently
supported by the study of Reguera and Rubi \cite{Reguera2001}. The
main idea is to combine all marginal higher orders corrections
$p_n(x)$ into a one-dimensional function $D(x)$, effectively acting
on the marginal probability density $p(x)$. In
Ref.~\cite{Reguera2001} the corrected stationary FJ-equation has the
form
\begin{align}
 0=\,-\partial_x J^{x}(x)=\,\p{}{x}\,D\bracket{x} e^{-A\bracket{x}} \p{}{x} e^{A\bracket{x}}
 p\bracket{x}\,, \label{eq:FJ_pdf_spatialD}
\end{align}
and was derived therein within the framework of mesoscopic
non-equilibrium thermodynamics. Kalinay and Percus
\cite{Kalinay2006} used a rigorous mapping of the $2$D diffusion
equation onto the reduced dimension and derived an expansion of the
diffusion coefficient $D(x)$, which represents corrections to the
FJ-equation.

In this spirit  we now determine the spatial dependent diffusion
coefficient $D(x)$ based on the presented results for the
perturbation series expansion \eq{eq:finalsol_pn}.  We concentrate
on the limit of small force strengths $|f| \ll 1$, so that diffusion
is the dominating process. Integrating the $2$D stationary
Smoluchowski equation \eq{eq:pdf_xy} over the cross-section in $y$,
and respecting the no-flux boundary condition \eq{eq:bc2}, one
derives an alternative definition of the marginal probability
current $J^{x}(x)$, equivalent to \eq{eq:FJ_pdf_spatialD}:
\begin{align}
 -J^x(x)=D\bracket{x}
 h(x)\partial_x\bracket{\frac{p\bracket{x}}{h(x)}}=
\intl{-h(x)}{h(x)}\partial_x p(x,y)dy\,. \label{eq:dspatial_1}
\end{align}
The second equality  determines the sought-after spatial dependent
diffusion coefficient $D(x)$. Note, that $D(x)$ is solely determined
by derivatives of $p(x,y)$ and $p(x)$. Hence, it plays no role
whether one uses the original expansion terms defined by
\eq{eq:longwave_pn} or the centered ones, given by
\eq{eq:def_shiftpn}. In compliance with Ref.~\cite{Kalinay2006}, we
make the ansatz that all but the first derivative of the boundary
function $h(x)$ are negligible. Then, the integration constants
$d_{n,1}(x)$ equal $0$ as they can been shown to be proportional to
higher derivatives of $h(x)$. Moreover, in the limit $|f|\ll 1$, the
$n$-times applied operator $\mathfrak{L}$, cf. \eq{eq:Ln},
simplifies to $ \mathfrak{L}^{n}=\,(-1)^n\,\p{ ^{2\,n}}{x^{2\,n}}$.
Moreover
\begin{align}
\mathfrak{L}^n p_0=\av{\dot{x}}_0(-1)^n
(2n-1)!\frac{(h')^{2n-1}}{2\, h^{2n}}+O(h''(x))\,.
\end{align}
Inserting the probability densities into \eq{eq:dspatial_1}, one
finds that
\begin{align}
 D(x)= &\sum_{n=0}^\infty \e^{2n}
 \bracket{-1}^n\,\frac{(h')^{2n}}{2n+1}+O(h''(x))\notag \\
 \simeq&\,\frac{\arctan\bracket{\varepsilon h'(x)}}{\varepsilon h'(x)} \label{eq:deff_kalinay_series}
\end{align}
for the spatially dependent diffusion coefficient $D(x)$ in the
diffusion dominated regime, i.e. when $|f|\ll 1$. Note, that this
expression for $D(x)$ was obtained previously by Kalinay and
Percus \cite{Kalinay2006} within a quite different expansion
approach.

In what follows, we evaluate  the average particle current
\eq{eq:series_vx_simp} by means of the spatially dependent diffusion
coefficient $D(x)$. According to $\lim_{f\to 0}
\av{\dot{x}}=\int_{0}^{1}dx\,J^x(x)$, it follows that in the small
force limit the mean particle velocity is proportional to the
expectation value of the spatially dependent diffusion coefficient,
yielding the main finding
\begin{align}
\begin{split}
 \lim_{f\to 0}\av{\dot{x}(f)} = &\,\lim_{f\to 0}\av{\dot{x}(f)}_0
 \av{D(x)}_x+O\bracket{h''(x)}  \\ \simeq&\lim_{f\to 0}\,\av{\dot{x}(f)}_0
 \av{\frac{\arctan\bracket{\varepsilon h'(x)}}{\varepsilon h'(x)}}_x\,.
\end{split}
\label{eq:mob_limf0}
\end{align}

In \eq{eq:mob_limf0}, the average of an arbitrary function $k(x)$ is
defined as the integral over one period divided by the period length
which is one in the considered scaling, i.e.,
$\av{k(x)}_x=\int_{0}^{1} k(x)\,dx$. In the linear response limit,
i.e. for $|f|\ll 1$, the Sutherland-Einstein relation emerges
Ref.~\cite{Burada2009_PTRSA,chaos05}, reading in dimensionless units:
\begin{align}
  \label{eq:SEinstein}
  \lim_{f\to0}\mu(f) = \lim_{f\to 0} D_{\mathrm{eff}}(f),
\end{align}
the effective diffusion coefficient $D_\mathrm{eff}$ is determined
by the mobility $\mu = \lim_{f\to 0}\av{\dot{x}(f)} /f$.
Consequently, if the average current $\av{\dot{x}(f)}_0$ (or the
effective diffusion coefficient $D_\mathrm{eff}^0(f)$) are known in
the zeroth order, the higher order corrections to both quantities can be obtained according to \eq{eq:mob_limf0}.

\section{Application of the theory to a sinusoidally shaped channel}
\label{sec:part_sinuschannel}

In the following we validate the obtained analytic predictions
\eq{eq:mob_limf0} with precise numerical simulations concerning one single
point-like Brownian particle moving with a corrugated sinusoidally-shaped geometry
Ref.~\cite{Burada2007,Burada2008}. The dimensionless boundary function
$h(x)$ reads
\begin{align}
 h_\pm\bracket{x}=&\,\pm h(x)=\,\pm\frac{1}{4}\bracket{\frac{1+\delta}{1-\delta}+\sin\bracket{2\pi\,x}}\,, \label{eq:conf2}
\end{align}
and is illustrated in \figref{fig:channelprob}. Please note, that in
absence of the scaling each channel geometry is determined by the
period $L$, the maximum width $\Delta \Omega$, and the width at the
bottleneck $\Delta \omega$. Upon scaling all lengths are measured in
units of the period $L$. Consequently, the parameter $\delta \Omega$
denotes the ratio of the maximum width $\Delta \Omega$ and the period
$L$, viz., $\delta \Omega=\Delta \Omega/L$. Equivalently, it holds
that $\delta \omega=\Delta \omega/L$. Within this scaling, the period
of the channel equals one. 

\begin{figure}[t]
\includegraphics{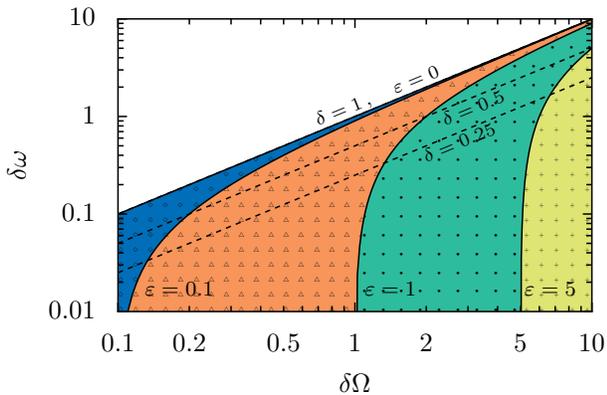}
\caption{(Color online) Schematic sketch of the dependence of the
  expansion parameter $\e=\delta\Omega-\delta\omega$ and the aspect
  ratio $\delta=\delta\omega/\delta\Omega$ on the maximum width
  $\delta \Omega$, respectively, the width at the bottleneck $\delta
  \omega$ in units of the period $L$. The dashed lines correspond to
  $\delta=1,0.5,0.25$ (from above) while the colored areas illustrate
  pairs of $(\delta\Omega,\delta \omega)$ where $\e\leq 0.1$
  (blue,circles), $\e\leq 1$ (red,triangles), $\e\leq 5$ (green,dots),
  and $\e>5$ (yellow,plus signs).} 
\label{fig:transf_parameter}
\end{figure}

In addition, one notices that the dimensionless boundary function
$h(x)$ is solely governed by the aspect ratio of the minimal and
maximal channel width $\delta=\delta \Omega/\delta \omega$. Obviously
different realizations of channel geometries can possess the same
value of $\delta$. The number of  orders have to taken into account in
the perturbation series \eq{eq:prob_series}, respectively, the
applicability of the {\it Fick-Jacob approach} to the problem, depends
only on the value of the slope parameter
$\e=\delta\Omega\bracket{1-\delta}$ for a given aspect ratio
$\delta$. For clarity, the impact of the maximum $\delta \Omega$ and
minimum width $\delta \omega$ on the expansion parameter $\e$,
respectively, the aspect ratio $\delta$ is illustrated in
\figref{fig:transf_parameter}.

According to the Sutherland-Einstein relation \eq{eq:SEinstein} the
mobility equals the effective diffusion coefficient (in the
dimensionless units) for $f\ll 1$ \cite{Burada2009_PTRSA}.
Consequently, it is sufficient to discuss the behavior of the
mobility $\mu(f)$. Referring to \secref{subsect:spatD}, the higher
order corrections to the mobility are given by the product of the
FJ-result and the expectation
value of the spatially dependent diffusion coefficient $D(x)$, see
\eq{eq:mob_limf0}.

First, we obtain the mobility $\mu_0$ within the zeroth-order
(Fick-Jacobs approximation). In the diffusion dominated regime, the
analytic expression for the mobility within the FJ-approach, cf. Eqs.~\eqref{eq:currFJ} and \eqref{eq:def_mob},
simplifies to the Lifson-Jackson formula \cite{Lifson1962,Burada2007}
\begin{align}
 \mu_{0} := \lim_{f\to 0} \mu_0(f)=&\, \frac{1}{\av{h(x)}\,\av{\frac{1}{h(x)}}} =\lim_{f\to0} D_\mathrm{eff}(f)\,.
\intertext{For the exemplarily considered channel geometry \eq{eq:conf2} the mobility attains the asymptotic value}
\lim_{f\to 0} \mu_0(f)=&\,\frac{2 \sqrt{\delta}}{1+\delta}=\frac{2\sqrt{1-\e/\delta \Omega}}{2-\e/\delta \Omega}\,. \label{eq:mobperi_FJ_limf0}
\end{align}
One notices that in the diffusion dominated regime $|f|\ll 1$ the
mobility of one single particle is determined only by the geometry -
more precisely by the aspect ratio $\delta$. In the limit of vanishing
bottleneck width, i.e. $\delta \to 0$,
the mobility tends to $0$. In contrast, for straight channels
corresponding to $\delta=1$, i.e. $\e = 0$, the
mobility equals its free value which is one in the considered scaling.

\begin{figure}[t]
\includegraphics{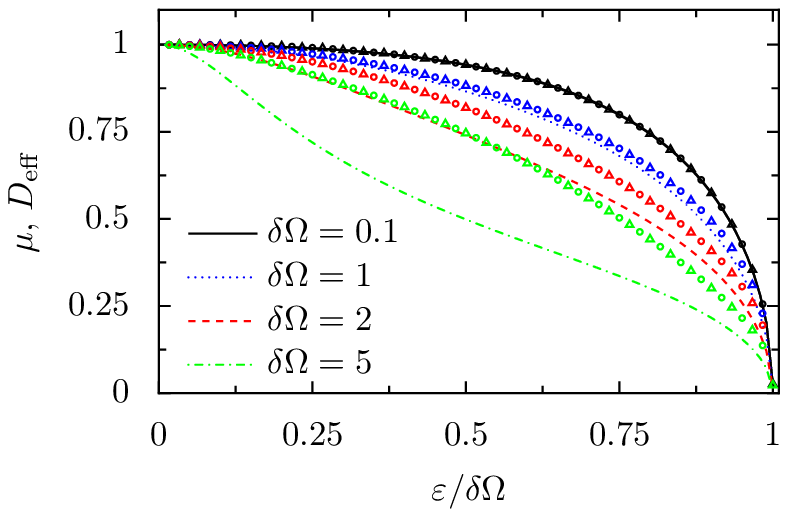}
\caption{(Color online) Comparison of the analytic theory versus precise
  numerics (in dimensionless units): The mobility and the effective diffusion
  constant for a Brownian particle moving inside a channel confinement
  are depicted as function of the ratio of slope parameter $\e$ and
  maximal channel width $\delta \Omega$ for different values $\delta\Omega = 0.1,1,2,5$ and bias $f=10^{-3}$  (corresponding to the diffusion dominated
  regime). The symbols correspond to the numerical obtained mobility (triangles) and the
  effective diffusion coefficient (circles).  The lines correspond to
  analytic higher order result, cf.~\eq{eq:mobperi_limf0}. The zeroth
  order - Fick-Jacobs results given by \eq{eq:mobperi_FJ_limf0}
  collapse to a single curve hidden by the solid line. }
\label{fig:mobdeffperi_eps}
\end{figure}

Evaluating  the period-averaged value of $D(x)$, i.e., considering
all higher order corrections apart from than scaling with higher derivatives of the boundary function
$h(x)$, we obtain from \eq{eq:mob_limf0}:
\begin{align}
  \mu :=& \lim_{f\to0} \mu(f) = \mu_{0} \, \av{D(x)}\nonumber \\ =&\,\,\frac{4\sqrt{1-\e/\delta\Omega}}{2-\e/\delta\Omega}\,\frac{\asinh\bracket{\pi \e /2}}{\pi\,\e} \label{eq:mobperi_limf0}
\end{align}
for the mobility $\mu$ and the effective diffusion coefficient $D_{\mathrm{eff}}$ in units of
its free values, respectively.

In \figref{fig:mobdeffperi_eps}, we depict the dependence of the $\mu(f)$ (triangles) and
$D_\mathrm{eff}(f)$ (circles) on the slope parameter $\e$
for $f=10^{-3}$. The numerical results are obtained by
solving the stationary Smoluchowski equation \eq{eq:Smoluchowski}
using finite element method \cite{FreeFem} and subsequently
calculating the average particle current \eq{eq:velocity}. In order to
determined the effective diffusion coefficient $D_\mathrm{eff}(f)$,
one has to solve numerically the reaction-diffusion equation for the {\itshape
B-field} \cite{Brenner,Laachi2007}. Note, that the numerical results for the effective diffusion coefficient $D_{\mathrm{eff}}(f)$ and the mobility $\mu(f)$ coincide for all values of $\e$, thus corroborating the Sutherland-Einstein relation. In addition, the Fick-Jacobs result, given by
\eq{eq:mobperi_FJ_limf0}, and the higher order result, see \eq{eq:mobperi_limf0},
are depicted in \figref{fig:mobdeffperi_eps}.

For the case of smoothly varying channel geometry, i.e. $\delta \Omega
\ll 1$, the analytic expressions are in excellent agreement
with the numerics, indicating the applicability of the Fick-Jacobs
approach. As long as the extension of the bulges of the channel structures is small compared
to the periodicity, sufficiently fast  transversal equilibration, which serves as
fundamental ingredient for the validity of the Fick-Jacobs
approximation is taking place. In virtue of \eq{eq:def_eps}, the slope parameter is defined by $\e=\delta \Omega-\delta \omega$ and hence the maximal value of $\e$ equals $\delta \Omega$, see \figref{fig:transf_parameter}. Consequently the influence of the higher expansion orders $\e^{2n}\av{\partial_x p_n(x,y)}$ on the average velocity \eq{eq:series_vx_simp} and mobility, respectively, becomes negligible if the maximum channel's width $\delta \Omega$ is small.

With increasing maximum width the difference between the
FJ-result  and the numerics is growing. Specifically, the FJ-approximation
resulting in \eq{eq:mobperi_FJ_limf0} overestimates the mobility $\mu$
and the effective
diffusion coefficient $D_{\mathrm{eff}}$. The higher order corrections
need to be included and consequently provide a good agreement for a wide range of
$\e$-values for maximum widths $\delta \Omega$ on the scale to the length of
the channel, i.e. $\delta \Omega\sim 1$, see the dotted line in \figref{fig:mobdeffperi_eps}.
Upon further increasing the maximum width $\delta \Omega$ diminishes
the range of applicability of the derived higher order corrections. This
is due to the neglect of the higher derivatives of the boundary
function $h(x)$. Put differently, the higher derivatives of $h(x)$ become significant for $\delta \Omega \gg 1$.

\section{Summary and conclusion}
\label{sec:conclusion}

In summary, we have considered the transport of point-size Brownian particles under the
influence of a constant and uniform force field through a
three-dimensional channel. The latter exhibits a constant height and
periodically varying side walls.

We have presented a systematic treatment of particle transport by
using a series expansion of the stationary probability density in terms of a smallness parameter which
specifies the corrugation of the channel walls. In particular, it
turns out that the leading order term of the series expansion is
equivalent to the well-established {\itshape Fick-Jacobs approach}
\cite{Jacobs,Zwanzig1992}. The higher order corrections to the
probability density become significant for extreme bending of the
channel's side walls. Analytic results for each order of the perturbation
series have been derived. Interestingly, within the 
presented perturbation theory, all higher
order corrections to the stationary probability
distribution and the average particle current scale with the average
particle current obtained from the Fick-Jacobs formalism. 
Moreover, by using the higher order
corrections we have derived an expression for
the spatially dependent diffusion coefficient $D(x)$ which
substitutes the constant diffusion coefficient present in the common
Fick-Jacobs equation. Accordingly, in the linear response regime,
i.e. for small forcing $|f|\ll 1$, the mean particle velocity is then
given by the product of the average particle current obtained from the
Fick-Jacobs formalism and the expectation value of the spatially
dependent diffusion coefficient $D(x)$ . Moreover, due to the
Sutherland-Einstein relation, the above statement also holds good for
the effective diffusion coefficient.   

Finally, we have applied our analytic results to a specific
example, namely, the particle transport through a channel with
sinusoidally varying side walls. We corroborate our theoretical
predictions for the mobility and the effective diffusion coefficient
with precise numerical results of a finite element calculation of the
stationary Smoluchowski-equation. In conclusion, the consideration of the
higher order corrections leads to a substantial improvement of the
Fick-Jacobs-approach, which corresponds to the zeroth order in our
perturbation analysis, towards more winding side walls of the
channel. 

\acknowledgments
\noindent This work has been supported by the VW
Foundation via project I/83903 (L.S.-G., S.M.) and I/83902 (P.H.,
G.S.). P.H. acknowledges the support by the DFG via SPP 1243,
the excellence cluster ''Nanosystems Initiative Munich'' (NIM), and
the German-Israeli Foundation (GIF, grant no. I 865-43.5/2005).

\end{document}